# Design and Modification of MEMS Based Micro Cantilever

Tejas S. Fanse[1]

[1](Department of Mechanical Engineering, Texas A&M University, Kingsville, Tx, USA)

**ABSTRACT** *: It goes without saying that at present MEMS Technology became one of the latest and emerging methods because of its miniaturization and effective cost. Now near the beginning recognition of disease became a major challenge in front of Researchers as well as Doctors. Thus, in this paper our main focus is to review about all these problems and give some solution. In order to do so here with proposed a micro cantilever-based sensor using MEMS technology in COMSOL Multiphysics environment using FEM. Analysis of Micro Cantilever sensor and its mechanical behavior as well as changing properties by changing in few parameters it has been observed that sensitivity and its deflection can be utilize to detect the different disease molecules and helpful for early detection. In this way it can be a helpful tool in the field of medical science.*

## I. INTRODUCTION

MEMS (Micro-Electro-Mechanical Systems) is a driver for multiple and mixed (materials, electronic, mechanical) technology integration. An emerging and one of the very strong technique MEMS is a device where microprocessors and mechanical parts along with signal processing circuits are integrated on a small piece of silicon.MEMS primary and a very unique feature is miniaturization, multiplicity as well as microelectronics (Sensors and actuators)[1-5].

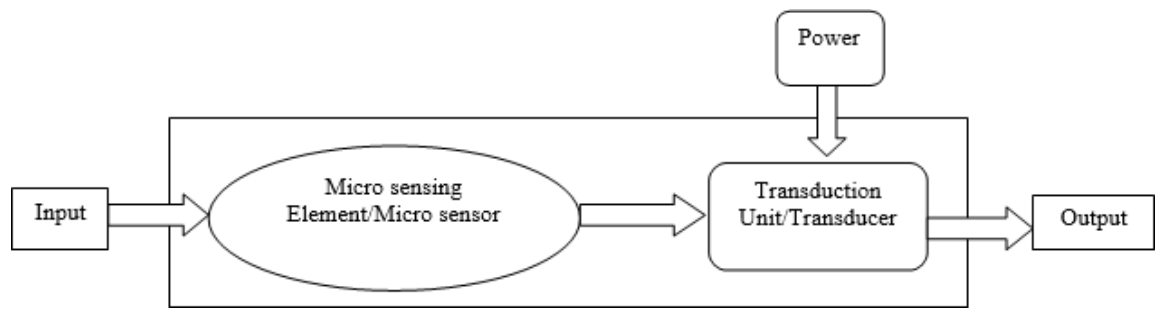

I.  Fig.-1 Basic Block Diagram of MEMS Operation

The enlargement of silicon technology has provided a number of significant rewards. Silicon is a tremendously good mechanical material. The micromechanical components can be integrated with the electronics to develop smart sensor and actuator systems with additional features such as self-test and self-calibration [8-14].

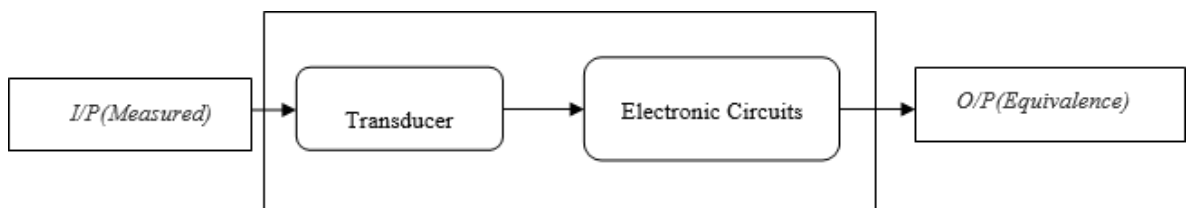

II. Fig.-2 Functionality of Transducer using Electronic Circuits

Sensors, actuators, electronics, computation, communication, control, power generation, chemical processing, biological reactions and many more things can be integrated, on a chip or in a package using Microsystems [17].The cantilever is one of the most famous and widely used structures in the field of microelectromechanical systems (MEMS) and Microsystems devices. Because of its flexibility and versatility popularity is very high in the field of MEMS based research [18-21].

It is a rigid beam or bar that is fixed to a support generally a vertical structure or wall and the beam's other end is free. Because of this horizontal beam that is firm at only one end while the other end is left free to hold some vertical loads. The beam's fixed end has a reaction force and moment created by the load acting at the free end. The purpose of a cantilever beam is to produce a bending outcome to a certain limit. These cantilevers are usually fabricated from silicon (Si), silicon nitride ($Si_3N_4$) or polymers are commonly made as unimorphs or bimorphs. There are so many possible shapes for micro cantilever-like Rectangular, Paddle shaped, triangular, trapezoidal, V-shaped, step profile, I-shaped, T-shaped and many more [3,10,11]. Since it can be shaped into the different structure using different materials so it's enhancing the uniqueness of microcantilever because of this different kind of diseases can be easily detected [3,4]. Cantilever mainly used as Biosensor and chemical sensor to detect many diseases with micro samples [1,6-12,19].

Cantilevers have some strong characteristics like very sensitive, fast measurement of mechanical movement and less power consumption. It can be operated in static mode or dynamic mode based on requirements it may use [6- 12]. The microcantilever is a broadly used component in MEMS (microelectromechanical systems). Because of its flexibility and adaptability, it is very popular for various applications. Cantilevers are available in all sizes. Microcantilevers range in length from a few meters to hundreds of meters. Microcantilevers are a few micrometers to several hundred micrometers in length. There are so many uses of Microcantilevers it may be used as sensors, transducers, probes, needles, transport mechanisms and switches for several tasks [20]. We may consider few of the examples a) Detect physical, chemical, and biological particles b) Penetrate tissue in therapeutic and diagnostic applications c) Sensors to detect nano-size particles on a surfaced) Memory storage devices. Microbeams and cantilever structures are basically very useful transducer elements using which a lot of physical phenomena can be measured. The principle behind this lies with the deflection of the beam and cantilever structures [2,5]. The deflection is picked up either by capacitive or piezoresistive measurement. The difference between a beam and cantilever is that the beam is fixed at both the ends whereas a cantilever is fixed at only one end. Some of the useful mathematical analysis is given as:

Mathematically, $\sigma = \dfrac{P}{A} ; N/mm^2$

Where, σ is stress is external load and A is Crossectional area.
1 KPa = $10^{-3}$ N/mm$^2$
1 MPa = 1N/mm$^2$
1 GPa = $10^3$ N/mm$^2$

Mathematically, $e = \dfrac{dl}{l}$

Where $(dl)$ change in length is, $l$ is original length and e is strain. It is unit less.

## II.SIMULATION AND DESIGN

In this paper using the FEM method in COMSOL Multiphysics environment microcantilever has been designed. As a shape rectangular structure has been chosen and size has been decided using following dimension L=500[µ m], W=50[µm] and Hight=10[µ m] apart from that most importantly the major part is the use of different materials. Here using Silicon Nitride, Silicon, Polysilicon, Silicon dioxide, Gold, Polyimide material microcantilever have been constructed. Since different materials have unlike properties so few other parameters are also considered which is given below in this Table 1.

Using the above parameters designed the following micro cantilever structures and its deflections can also be seen in these pictures along with its Eigen frequency. Frequencies at which a system is prone to vibrate is called Eigen frequencies or natural frequencies, these frequencies are discrete in nature. In this paper, an effort is made to study the behaviour of Eigen frequencies with the help of mechanical structures cantilever. In this Eigen frequency analysis, the only shape of the mode, not the amplitude of any physical vibration can be analysed. Herewith it has been observed that periodic excitation does not cause a resonance that may lead to excessive stresses or noise emission simultaneously it may Check if a quasistatic analysis of a structure is appropriate based on the fact that all natural frequencies are high when compared to the frequency content of the loading and examine appropriate choices of time steps or frequencies for a subsequent dynamic response analysis. Fig- 4 to Fig.9 showing the deflection and Eigen frequencies of micro cantilever while changing the materials as Silicon Nitride, Silicon, Polysilicon, Silicon dioxide, Gold, Polyimide respectively.

In the above figures only one eigenfrequency has been shown using six different materials but for more clarity the table which is given below showing six different frequencies in each and every case.

### III. FIGURES AND TABLES

| Property | Variable | Each material have different Values (Used materials) | Unit | Property |
|---|---|---|---|---|
| Density | rho | Silicon Nitride, Silicon, Polysilicon, Silicon dioxide, Gold, Polyimide | Kg/m$^3$ | Basic |
| Young's Modulus | E | | pa | Young's Modulus & Poisson's Ratio |
| Poisson's Ratio | nu | | 1 | Young's Modulus & Poisson's Ratio |

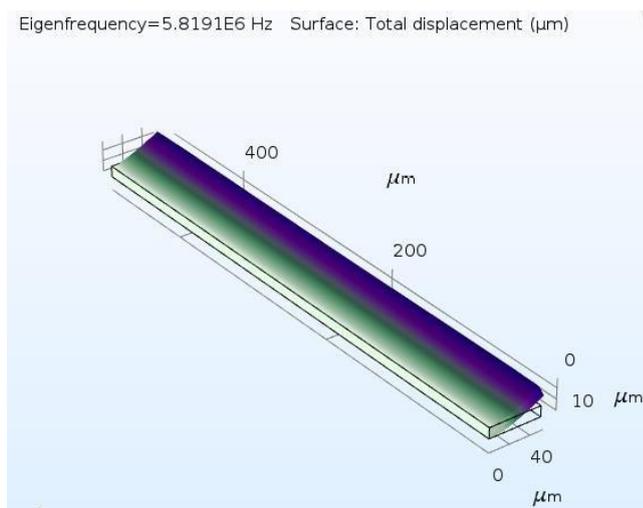 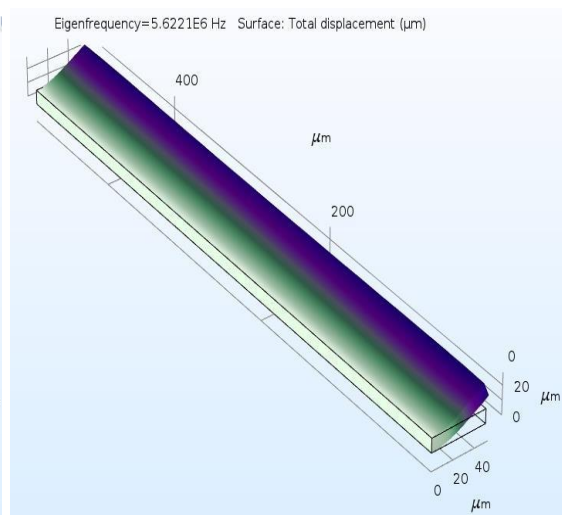

I. FIG.4 EIGEN FREQUENCY USING SILICON NITRIDE      FIG.5 EIGEN FREQUENCY USING SILICON

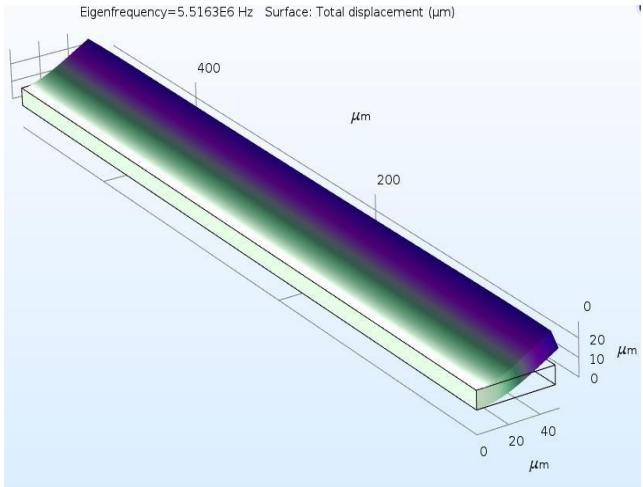

II. FIG.6 EIGEN FREQUENCY USING POLYSILICON

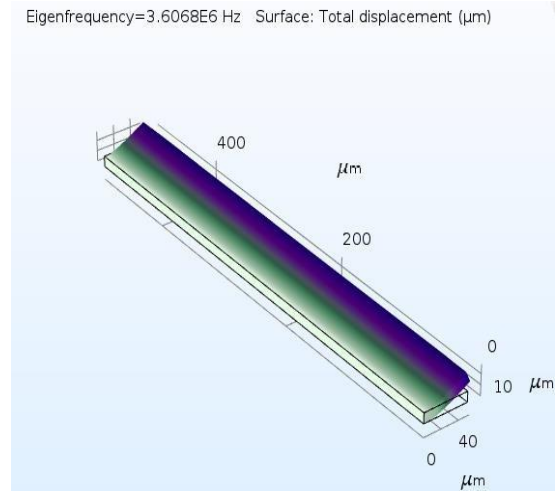

FIG.7 EIGEN FREQUENCY USING SILICON DIOXIDE

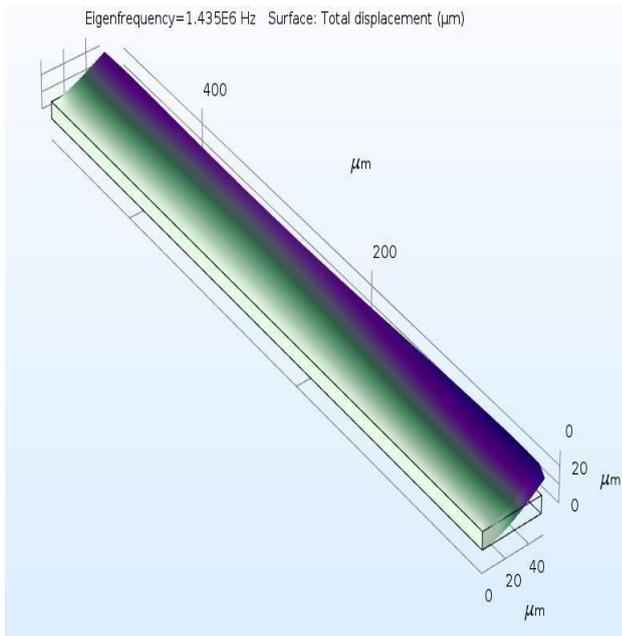

Fig.8 Eigen Frequency using Gold

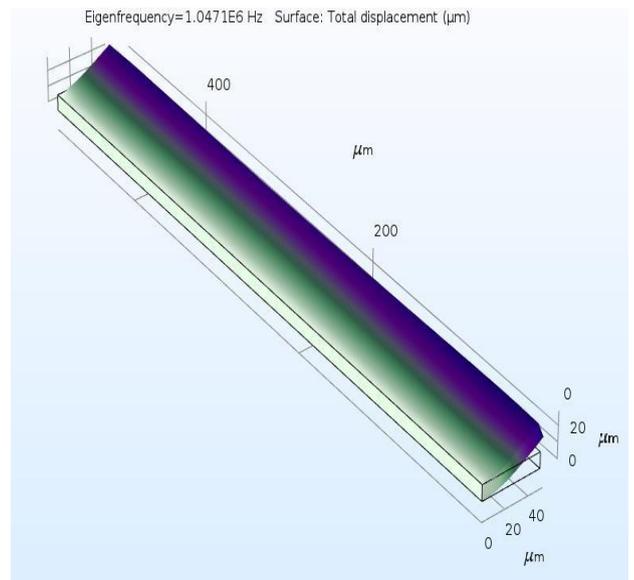

Fig.9 Eigen Frequency using Polyimide

In the above figures only one eigenfrequency has been shown using six different materials but for more clarity the table which is given below showing six different frequencies in each and every case.

The Table 2 and 3 given below shows the different Eigenfrequencies while changing the cantilever materials. In the field of medical science, microcantilever may utilize based on their eigenfrequencies using different methods like piezoelectric, piezoresistive, optical or electro statistics.

III. TABLE-2 EIGEN FREQUENCIES USING DIFFERENT MATERIALS

| Dimensions (μm) | Eigen Frequencies | | | | | |
|---|---|---|---|---|---|---|
| | Silicon Nitride | Silicon | Polysilicon | Silicon Oxide | Gold | Polyimide |
| L=500(μm) W=50(μm) H=10(μm) | 5.8191E6 | 5.6221E6 | 5.5163E6 | 3.6068E6 | 1.435E6 | 1.0471E6 |
| | 5.9507E6 | 5.7351E6 | 5.6437E6 | 3.6985E6 | 1.4502E6 | 1.0642E6 |
| | 6.3626E6 | 6.1038E6 | 6.0398E6 | 3.9759E6 | 1.5198E6 | 1.1251E6 |
| | 7.0366E6 | 6.7127E6 | 6.6867E6 | 4.4241E6 | 1.6322E6 | 1.2265E6 |
| | 7.9676E6 | 7.5666E6 | 7.5777E6 | 5.0333E6 | 1.8027E6 | 1.3723E6 |
| | 9.1495E6 | 8.6641E6 | 8.7063E6 | 5.7962E6 | 2.0323E6 | 1.5632E6 |

**Table-2 comparisons and applications**

| Dimensions (μm) | Highest, Lowest Eigen Frequencies and Medical Applications | | | |
|---|---|---|---|---|
| | Material | Highest Eigen Frequency | Lowest Eigen Frequency | Medical Applications |
| L=500(μm) W=50(μm) H=10(μm) | Silicon Nitride | 9.1495E6 | 5.8191E6 | Pathogen |
| | Silicon | 8.6641E6 | 5.6221E6 | Liver cancer |
| | Polysilicon | 8.7063E6 | 5.5163E6 | Glucose Sensing |
| | Silicon Oxide | 5.7962E6 | 3.6068E6 | HIV virus |
| | Gold | 2.0323E6 | 1.435E6 | Cancer |
| | Polyimide | 1.5632E6 | 1.0471E6 | Medical Equipments |

Through the above investigation it is found that Silicon Nitride has greater Eigen frequencies as compare to other materials.

## IV. CONCLUSION

This COMSOL environment-based simulation analysis shows the importance of dimensions and materials. Depends on changing the materials Eigen frequencies or natural frequencies will also be varied. At last, it has been observed that Silicon Nitride has greater Eigen frequencies as compare to other materials. So using different methods like piezoelectric, piezoresistive, optical or electro statistics micro cantilever can be used in the field of medical science for the identification of various diseases.